\documentclass{appolb}
\usepackage{graphicx}
\usepackage{amssymb,amsmath,amsfonts}
\usepackage[utf8]{inputenc}
\usepackage[normalem]{ulem}
\usepackage{color}
\usepackage{xspace}

\newcommand{\ud}{\mathrm{d}}

\newcommand{\mub}{\ensuremath{\mu_{B}}\xspace}
\newcommand{\dndy}{\ensuremath{\ud N/\ud y}\xspace}

\renewcommand\sout{\bgroup \color{blue} \ULdepth=-.5ex \ULset}

\begin{document}

\title{Statistical hadronization: successes and some open issues}

\author{Anton Andronic
\address{Institut f\"ur Kernphysik, Universit\"at M\"unster, 48149  M\"unster, Germany}
\\
Peter Braun-Munzinger
\address{Research Division and EMMI, GSI Helmholtzzentrum f\"ur Schwerionenforschung, 64291 Darmstadt, Germany}
\address{Physikalisches Institut, Universit\"at Heidelberg, 69120 Heidelberg, Germany}
\address{Institute of Particle Physics and Key Laboratory of Quark and Lepton Physics (MOE), Central China Normal University, Wuhan 430079, China}
\\
Krzysztof  Redlich
\address{University of Wroc\l aw, Institute of Theoretical Physics, 50-204 Wroc\l aw, Poland}
\\
Johanna  Stachel
\address{Physikalisches Institut, University of Heidelberg, 69120 Heidelberg, Germany}
\address{Institute of Particle Physics and Key Laboratory of Quark and Lepton Physics (MOE), Central China Normal University, Wuhan 430079, China}
}

\maketitle 

\begin{abstract}
Hadron production in relativistic nuclear collisions is well described in the framework of the statistical hadronization model, over a broad range of collision energies. We outline this for hadrons composed of light (u, d, s) and heavy (charm and beauty) quarks, discuss recent findings relevant for understanding the phase structure of QCD and formulate some open issues.
\end{abstract}

\section{Introduction} \label{sect:intro}

If one compresses or heats strongly interacting matter to higher densities and/or high temperatures one expects \cite{Itoh:1970uw,Collins:1974ky,Cabibbo:1975ig,Chapline:1976gy} that quarks are no longer confined but can move over distances significantly larger than the size of a nucleon. At very similar conditions, also a fundamental symmetry of QCD, the chiral symmetry that is spontaneously broken in hadronic and nuclear matter, is restored. 
Such a deconfined, chirally symmetric state of matter, the Quark-Gluon Plasma (QGP) \cite{Shuryak:1978ij}, is likely to have existed in the Early Universe between the electroweak phase transition at picoseconds after the Bib Bang and for up to 10 microseconds \cite{Boyanovsky:2006bf}. It can be studied experimentally and theoretically via collisions of nuclei at high energies \cite{Busza:2018rrf,Braun-Munzinger:2015hba}.
One stage in the complex dynamics of the system produced in heavy-ion collisions is that of the chemical freeze-out, at which the abundances of hadron species are fixed. Chemical freeze-out is addressed phenomenologically within the statistical hadronization model (SHM) \cite{BraunMunzinger:2003zd,Andronic:2005yp,Andronic:2017pug}.
The value of the pseudo-critical temperate $T_c$ for the chiral crossover transition at vanishing \mub is currently calculated in Lattice QCD (LQCD) to be 156.5$\pm 1.5$ MeV \cite{Bazavov:2018mes} and 158.0$\pm 0.6$ MeV \cite{Borsanyi:2020fev}.
LQCD results also quantify a small decrease of $T_c$ with increasing \mub as long as $\mub \lesssim 400$ MeV \cite{Bonati:2018nut,Bazavov:2018mes,Borsanyi:2020fev}. Within this parameter range the chiral transition is still of crossover type \cite{Aoki:2006we}. The temperature for the deconfinement transition is more difficult to evaluate in LQCD due to the lack of an order parameter at finite quark masses. A recent extrapolation of the static quark entropy \cite{Borsanyi:2024xrx} puts the deconfinement transition line very close to the one for the chiral transition. The presence of new phases and possibly of a critical end point at values of $\mu_B$ around 600 MeV is currently theoretically discussed intensely and addressed experimentally, see very recent reviews~\cite{Fischer:2026uni,Braun-Munzinger:2026krf}.

One of the consequences of confinement in QCD is that physical observables require a representation in terms of hadronic states. Indeed, as has been noted in the context of QCD thermodynamics (see, e.g., \cite{Bazavov:2017dus} and refs. therein) the corresponding partition function $Z$ can be very well approximated within the framework of the hadron resonance gas, as long as the temperature stays below $T_c$.

In a volume V the grand canonical partition function for  hadron species $i$ is: 
\begin{equation}
\ln Z_i ={{Vg_i}\over {2\pi^2}}\int_0^\infty \pm p^2\ud p \ln [1\pm  \exp (-(E_i-\mu_i)/T)]
\end{equation}
with $+$ for fermions and $-$ for bosons, where $g_i=(2J_i+1)$ is the spin degeneracy factor, $T$ is the temperature, $E_i =\sqrt {p^2+m_i^2}$ the total energy;
$\mu_i = \mu_B B_i+\mu_{I_3} I_{3i}+\mu_S S_i+\mu_C C_i$ are the chemical potentials that ensure conservation (on average) of baryon, isospin, strangeness, and charm quantum numbers.  Three initial conditions help fixing ($I_{3i},\mu_S,\mu_C$):
i) isospin stopping identical to baryon stopping: $I_{3}^{tot}/\sum_i n_i I_{3i}= N_B^{tot}/\sum_i n_i B_i$, with $I_{3}^{tot}$ and $N_{B}^{tot}$ as the isospin and baryon numbers of the system ;
ii) vanishing net initial strangeness: $\sum_i n_i S_i = 0$; iii) vanishing net initial charm content: $\sum_i n_i C_i = 0$.

One needs as input for the calculations knowledge of the complete hadron spectrum and the default constitutes what is listed by the PDG \cite{Zyla:2020zbs}.

A consistent approach to interactions among hadrons is an implementation employing the S-matrix formulation of statistical mechanics with measured pion-nucleon phase shifts including, importantly, also repulsive and non-resonant components \cite{Andronic:2018qqt}. This accounts for the strongest contribution, the pion-nucleon interaction. Implementation of interactions in the strangeness sector are realized in ~\cite{Cleymans:2020fsc}. For an interesting new contribution  see ~\cite{Yasui:2026vve}.

\section{Statistical hadronization of light quarks}

\begin{figure}[hbt]
\begin{tabular}{cc}  \begin{minipage}{.56\textwidth}
 \hspace{-.4cm}  \includegraphics[width=1.05\textwidth]{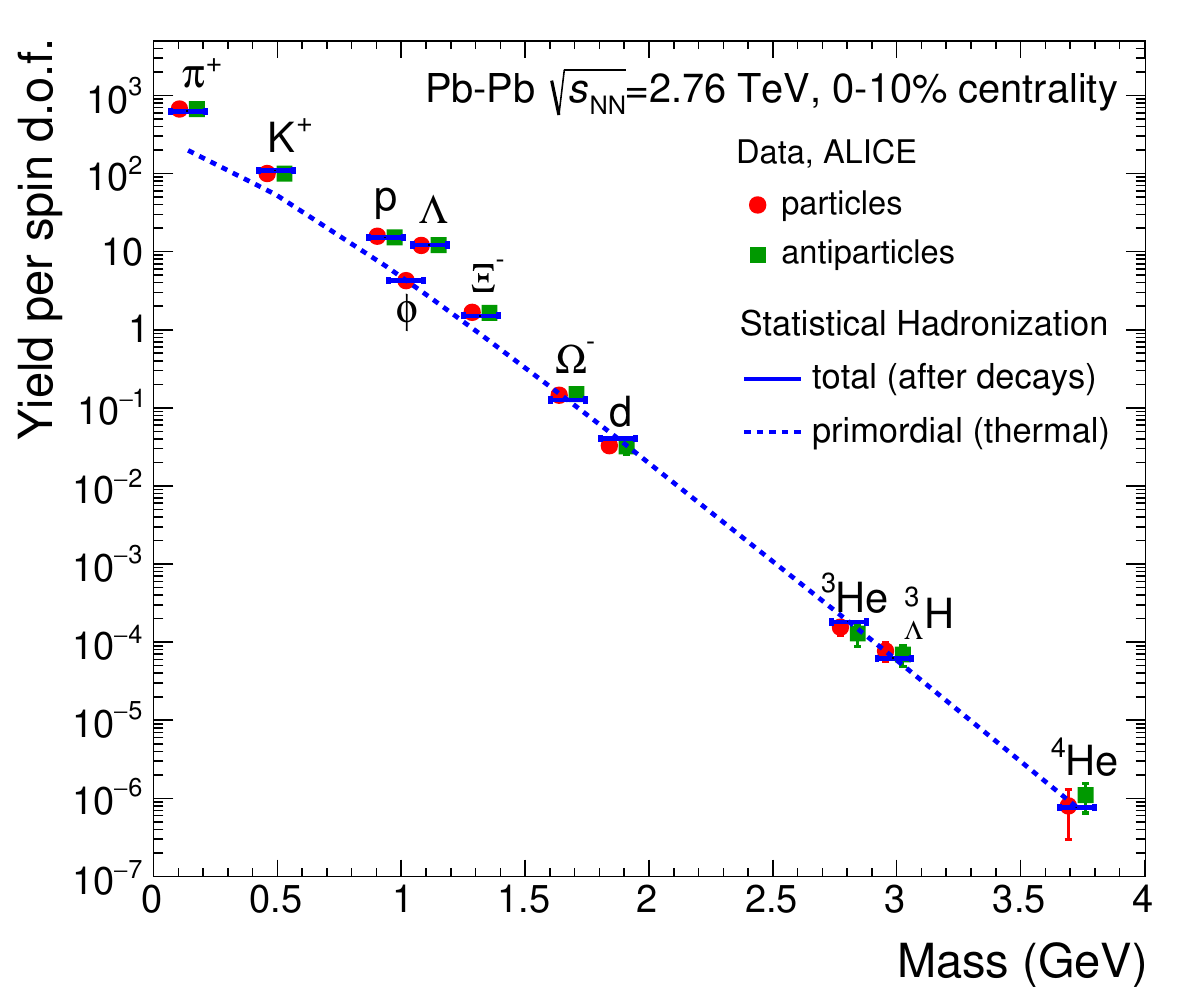}
\end{minipage} & \begin{minipage}{.4\textwidth}
\caption{Mass dependence of hadron yields divided by the spin degeneracy factor ($2J+1$), SHM best fit in comparison to the ALICE data. For the SHM are plotted both the ``total'' yields, including strong and electromagnetic decays, and the ``primordial'' thermal yields. Figure taken from~\cite{Andronic:2021dkw}.
}\label{fig:Fit}
\end{minipage} \end{tabular}
\end{figure}

In practice, $T_{CF}$, $\mub$, and $V$, the parameters at chemical freeze-out are determined from a fit to the experimental data. 
For the most-central (0-10\%) Pb--Pb collisions at the LHC, the best description of the ALICE data (see \cite{Acharya:2017bso} and ref. therein) on yields of particles in one unit of rapidity at midrapidity, is obtained with $T_{CF}=156.6\pm 1.7$ MeV, $\mu_B=0.7\pm 3.8$ MeV, and $V=4175\pm 380$ fm$^3$ (corresponding to a slice of one unit of rapidity, centered at mid-rapidity) \cite{Andronic:2017pug,Andronic:2018qqt}, shown in Fig.~\ref{fig:Fit}.
The standard deviations quoted here are exclusively due to experimental uncertainties and do not reflect the systematic uncertainties connected with the model implementation. Further investigations have led to an order of magnitude improvement in the precision of $\mub =0.7\pm0.45$ MeV ~\cite{ALICE:2023ulv}, demonstrating that the central region at LHC energies is essentially baryon-free.

Very good agreement is obtained between the measured particle yields and SHM over nine orders of magnitude in abundance values, encompassing strange and non-strange mesons, baryons, including strange and multiply-strange hyperons, as well as light nuclei and hypernuclei and their anti-particles.
The initially-observed over prediction of about 20\% of the data by the model for proton and anti-proton yields (a deviation of 2.7$\sigma$) is entirely accounted for via the S-matrix treatment of the interactions \cite{Andronic:2018qqt} leading to an excellent fit with a $\chi^2_{red} = 16.9/19$ (for consistency, with the S-matrix treatment the excluded-volume correction is not applied anymore). It was recently shown that the addition (compared to what is listed by PDG \cite{Zyla:2020zbs}) of about 500 new states predicted by LQCD and the quark model does lead to a strong deterioration of the fit, while a restoration of the good fit quality at no change of the thermal parameters is observed when the S-matrix treatment is employed as well for this expanded hadron spectrum \cite{Andronic:2020iyg}. 

The thermal origin of all particles including light nuclei and anti-nuclei is particularly transparent when inspecting the dependence of their yields with particle mass, shown in the right panel of Fig.~\ref{fig:Fit}.
We note that the yields of the measured lightest mesons and baryons, ($\pi,K,p,\Lambda$) are substantially increased relative to their primordial thermal production by the resonance decay contributions (for pions, e.g., the decay contribution amounts to 70\% of the total yield).
For the subset of light nuclei, the SHM predictions are, however, not affected by resonance decays.
For these nuclei, due to their large masses, a small variation in temperature leads to a large variation of the yield, resulting in a relatively precise determination of the freeze-out temperature $T_{nuclei} = 159 \pm 5$ MeV, well consistent with the value of $T_{CF}$ extracted above.

The rapidity densities of light (anti-)nuclei and hypernuclei were actually predicted \cite{Andronic:2010qu}, based on the systematics of hadron production at lower energies. It is nevertheless remarkable that such loosely bound objects (the deuteron binding energy is 2.2 MeV, much less than $T_{CF} \approx T_c  \approx 157$ MeV) are produced with temperatures very close to that of the phase boundary at LHC energy, implying any further evolution of the fireball has to be close to isentropic.
The detailed production mechanism for loosely bound states remains an open question (see recent review \cite{Braun-Munzinger:2018hat}). 
One possibility is that such objects, at QGP hadronization, are produced as compact, colorless droplets of quark matter with quantum numbers of the final state hadrons~\cite{Andronic:2017pug} (see discussion below).

\begin{figure}[htb]
  \includegraphics[width=.46\textwidth]{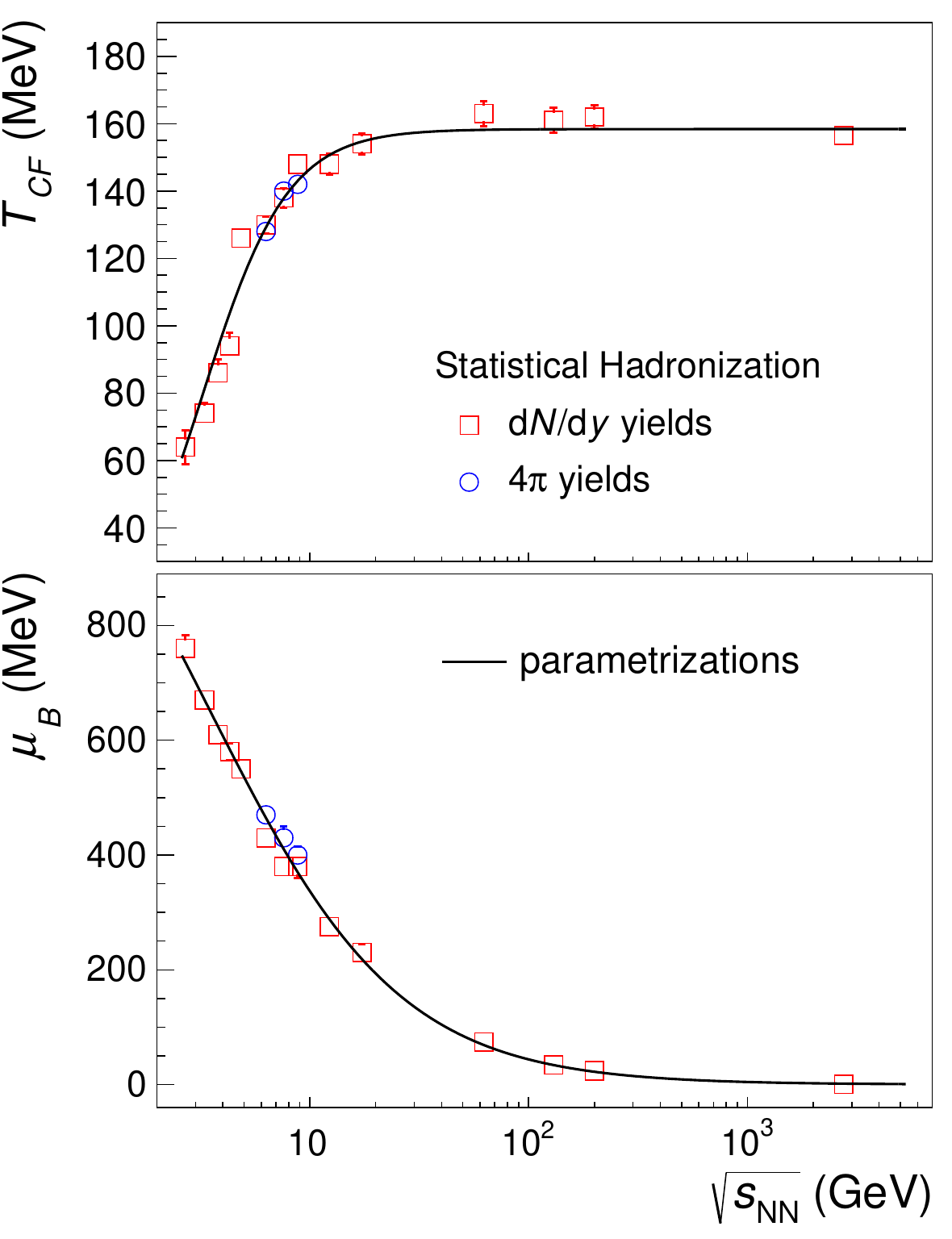} \includegraphics[width=.52\textwidth]{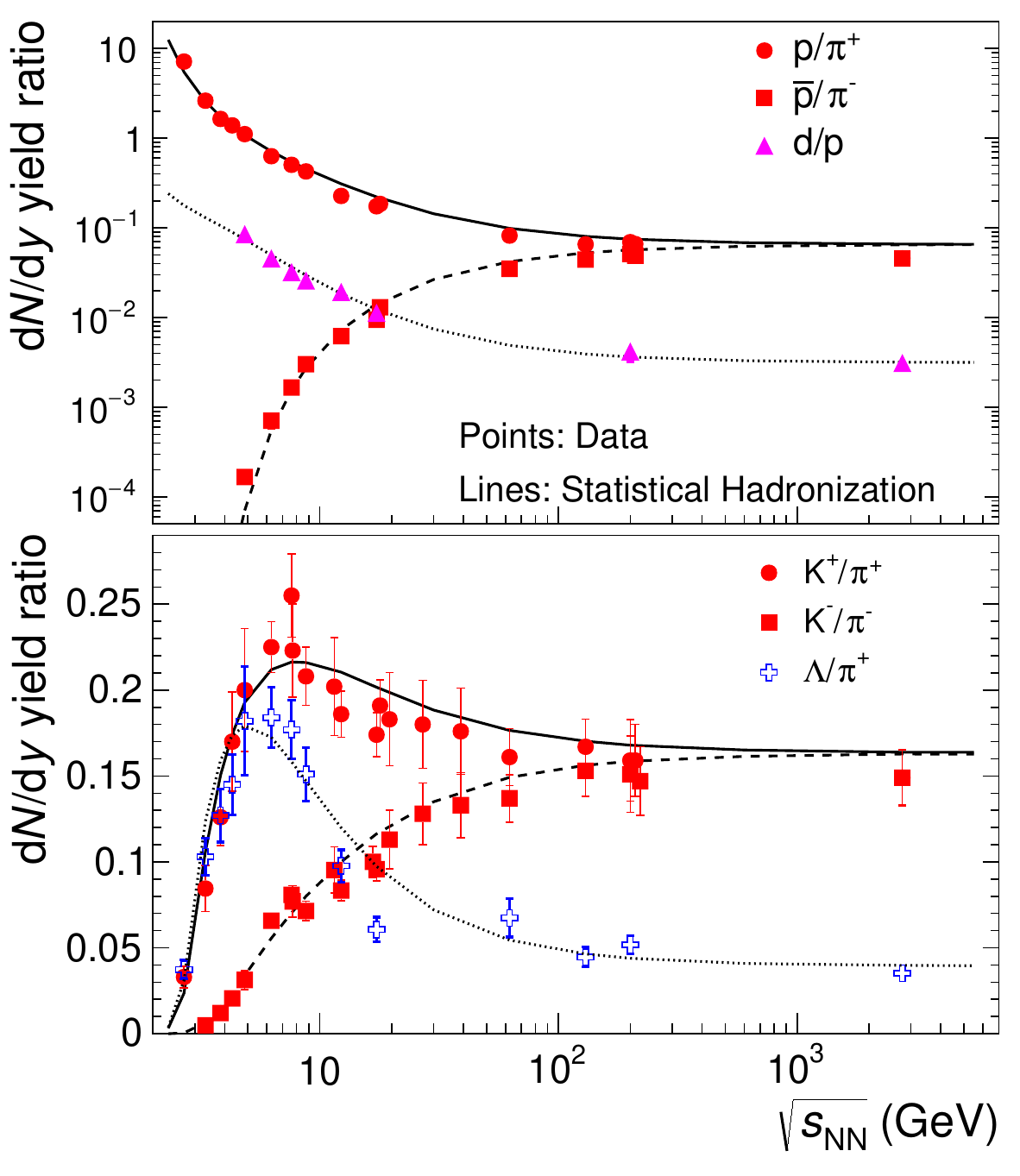}
\caption{Left: Energy dependence of chemical freeze-out parameters $T_{CF}$  and \mub. The results are obtained from the SHM analysis of hadron yields (at midrapidity, \dndy, and in full phase space,  $4\pi$) for central collisions at different energies. Right: Collision energy dependence of the relative abundance of several hadron species (the data are compiled in \cite{Andronic:2014zha,Adamczyk:2017iwn}). Figures taken from~\cite{Andronic:2017pug}.}
\label{fig:edep}
\end{figure}

The thermal nature of particle production in ultra-relativistic nuclear collisions has been experimentally verified not only at LHC energy, but also at the lower energies of the RHIC, SPS and AGS accelerators. The essential difference is that, at these lower energies, the matter anti-matter symmetry observed at the LHC is lifted, implying non-vanishing values of the chemical potentials. Furthermore, in central collisions at energies below $\sqrt{s_{\rm NN}} \approx 6 $ GeV the cross section for the production of strange hadrons decreases rapidly, with the result that the average strange hadron yields per collision can be significantly below unity. In this situation, one needs to implement exact strangeness conservation, applying canonical thermodynamics~\cite{Hagedorn:1984uy,Hamieh:2000tk,Braun-Munzinger:2003pwq}. Similar considerations apply for the description of particle yields in peripheral nuclear and elementary collisions.

While \mub decreases smoothly with increasing energy, the dependence of $T_{CF}$ on energy exhibits a striking feature which is illustrated in Fig.~\ref{fig:edep}: $T_{CF}$ increases with increasing energy from about 60 MeV to a saturation for $\sqrt{s_{\rm NN}} > 20$ GeV, of about 158 MeV, when averaging over all experiments.  
The saturation of $T_{CF}$ observed in Fig.~\ref{fig:edep} lends support to the earlier proposal \cite{BraunMunzinger:1998cg,Stock:1999hm,BraunMunzinger:2003zz} that, at least at high energies, the chemical freeze-out temperature is very close to the QCD hadronization temperature \cite{Andronic:2008gu}, implying a direct connection between data from relativistic nuclear collisions and the QCD phase boundary. Hagedorn noted long ago~\cite{Hagedorn:1965st} that hadronic matter cannot be heated beyond a certain limit, but the saturation observed here indicates a different boundary. The parametrizations shown in Fig.~\ref{fig:edep} are:
$T_{CF}={T_{CF}^{lim}}/(1+\exp(2.60-\ln(\sqrt{s_{\rm NN}})/0.45))$ and
$\mub ={a}/(1+0.288\sqrt{s_{\rm NN}})$, with $\sqrt{s_{\rm NN}}$ in GeV,
 the 'limiting temperature' $T_{CF}^{lim}=158.4\pm 1.4$ MeV, and $a=1307.5$ MeV. 

To illustrate how well the thermal description of particle production in central nuclear collisions works we show also in Fig.~\ref{fig:edep} (right), the energy dependence of the relative abundance of several hadron species along with the prediction using the SHM and the parametrized evolution of the parameters.
In particular, the maxima  (occurring at slightly different c.m. energies) in the $K^+/\pi^+$ and $\Lambda/\pi^+$ ratios are naturally explained \cite{Andronic:2008gu} as the interplay between the energy dependence of $T_{CF}$ and $\mu_B$ and the consequence of strangeness conservation.
Deuterons are also well reproduced (see discussion below).

\vspace{-3mm}
\begin{figure}[htb]
\begin{tabular}{cc}  \begin{minipage}{.48\textwidth}
 \hspace{-.4cm}   
 \includegraphics[width=1.04\textwidth]{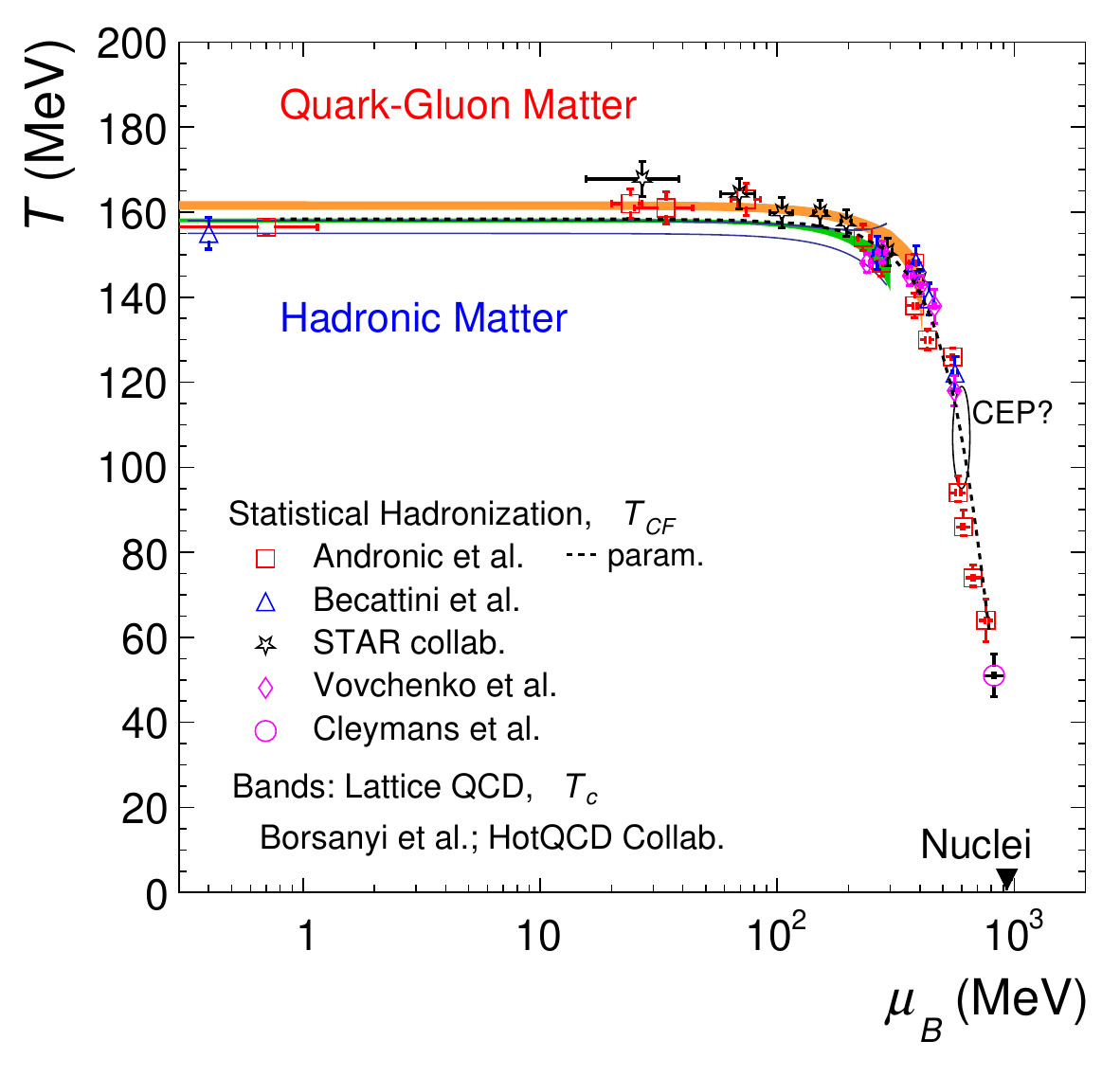}
\end{minipage} & \begin{minipage}{.47\textwidth}
\vspace{-7mm}
\caption{Phase diagram of strongly interacting matter constructed from chemical freeze-out points for central collisions at different energies, extracted from experimental data sets in our own (squares) and other similar analyses \cite{Cleymans:1998yb,Vovchenko:2015idt,Becattini:2016xct,Adamczyk:2017iwn} are compared to predictions from LQCD \cite{Bazavov:2018mes,Borsanyi:2020fev,Borsanyi:2024xrx} shown as bands. For the location of a possible critical endpoint (CEP) see~\cite{Fischer:2026uni}. 
}
\label{fig:t-mu}
\end{minipage}\end{tabular}
\end{figure}

Since the statistical hadronization analysis at each collision energy yields a pair of ($T_{CF}$,\mub) values, these points can be entered into the phase diagram of QCD, shown in Fig.~\ref{fig:t-mu}. Note that the points at low temperature and high $\mu_B$ seem to converge towards the value for ground state nuclear matter ($\mub\simeq930$~MeV). 
At high collision energies the points are very close to the pseudo-critical line for the chiral phase transition (and deconfinement)~\cite{Bazavov:2018mes,Borsanyi:2020fev,Borsanyi:2024xrx}.

\section{Extension to hadrons with heavy quarks}

There is now significant experimental information, from relativistic nuclear collisions, not only on the production of hadrons composed of light (u,d,s) valence quarks, but also of open and hidden charm and beauty hadrons. In particular, there is good evidence, mainly from results obtained at the CERN Large Hadron Collider (LHC) ~\cite{ALICE:2021rxa,Andronic:2021erx,Andronic:2019wva}, that charm quarks reach a large degree of thermal equilibrium, although charm quarks in the system are initially chemically far out of equilibrium. This is supported by heavy quark diffusion coefficients from LQCD~\cite{Altenkort:2020fgs,Altenkort:2023oms}. A strong indication for equilibration is the fact that J/$\psi$ mesons participate in the collective, anisotropic hydrodynamic expansion ~\cite{ALICE:2013xna,He:2021zej}.

To microscopically understand the production me\-cha\-nism of charmed hadrons for systems ranging from pp to Pb--Pb, various forms of quark coalescence mo\-dels have been developed ~\cite{Cho:2019lxb,ExHIC:2017smd,Zhou:2014kka,Greco:2003vf}. This provides a natural way to study the dependence of production yields on hadron size and, hence, may help to settle the still open question whether the many exotic hadrons that have been observed recently are compact multi-quark states or hadronic molecules (see ~\cite{Aarts:2016hap,Maiani:2022psl} and refs. cited there). Conceptual difficulties with this approach are that energy is not conserved in the coalescence process and that color neutralization at hadronization requires additional assumptions about quark correlations in the QGP~\cite{Song:2021mvc}.

Another approach, named SHMc, has been formulated by the extension of the SHM to also incorporate charm quarks. This was first proposed in ~\cite{Braun-Munzinger:2000csl} and developed further in ~\cite{Andronic:2003zv,Andronic:2006ky,Andronic:2017pug,Andronic:2021erx} to include all hadrons with hidden and open charm. The key idea is based on the recognition that, contrary to what happens in the (u,d,s) sector, the heavy (mass $\sim$ 1.2 GeV) charm quarks are even at LHC energy not thermally produced. Rather, production takes place in initial hard collisions. The produced charm quarks then thermalize in the hot fireball, but the total number of charm quarks is conserved during the evolution of the fireball~\cite{Andronic:2006ky} since charm quark annihilation is very small. In essence, this implies that charm quarks can be treated like impurities. Their thermal description then requires the introduction of a charm fugacity $g_c$~\cite{Braun-Munzinger:2000csl,Andronic:2021erx}. The value of $g_c$ is not a free parameter but experimentally determined by measurement of the total charm cross section. For central Pb--Pb collisions at LHC energy, $g_c \approx 30$~\cite{Andronic:2021erx}. The charmed hadrons are, in the SHMc, all formed at the phase boundary, i.e. at hadronization, in the same way as all (u,d,s) hadrons, only with the boundary condition that all charm quarks present in the QGP materialize in hadrons (as warranted by $g_c$).

In ~\cite{Andronic:2019wva} it is demonstrated that, with that choice, the measured yield for J/$\psi$ mesons is very well reproduced along with the yield of all light-flavor hadrons. The uncertainty in the prediction is mainly caused by the uncertainty in the total charm cross section in Pb--Pb collisions. We note here that the excellent agreement of charmed hadron yields with those computed with the SHMc implies that charm quarks, and consequently 
charmonia, are unbound inside the QGP; in fact their final yields at full LHC energy exhibit enhancement compared to expectations using collision scaling from pp collisions, contrary to the original predictions based on ~\cite{Matsui:1986dk}. For a detailed discussion see ~\cite{Andronic:2017pug}.

\begin{figure}[!htb]
    \centering
    \includegraphics[width=.52\linewidth,clip=true]{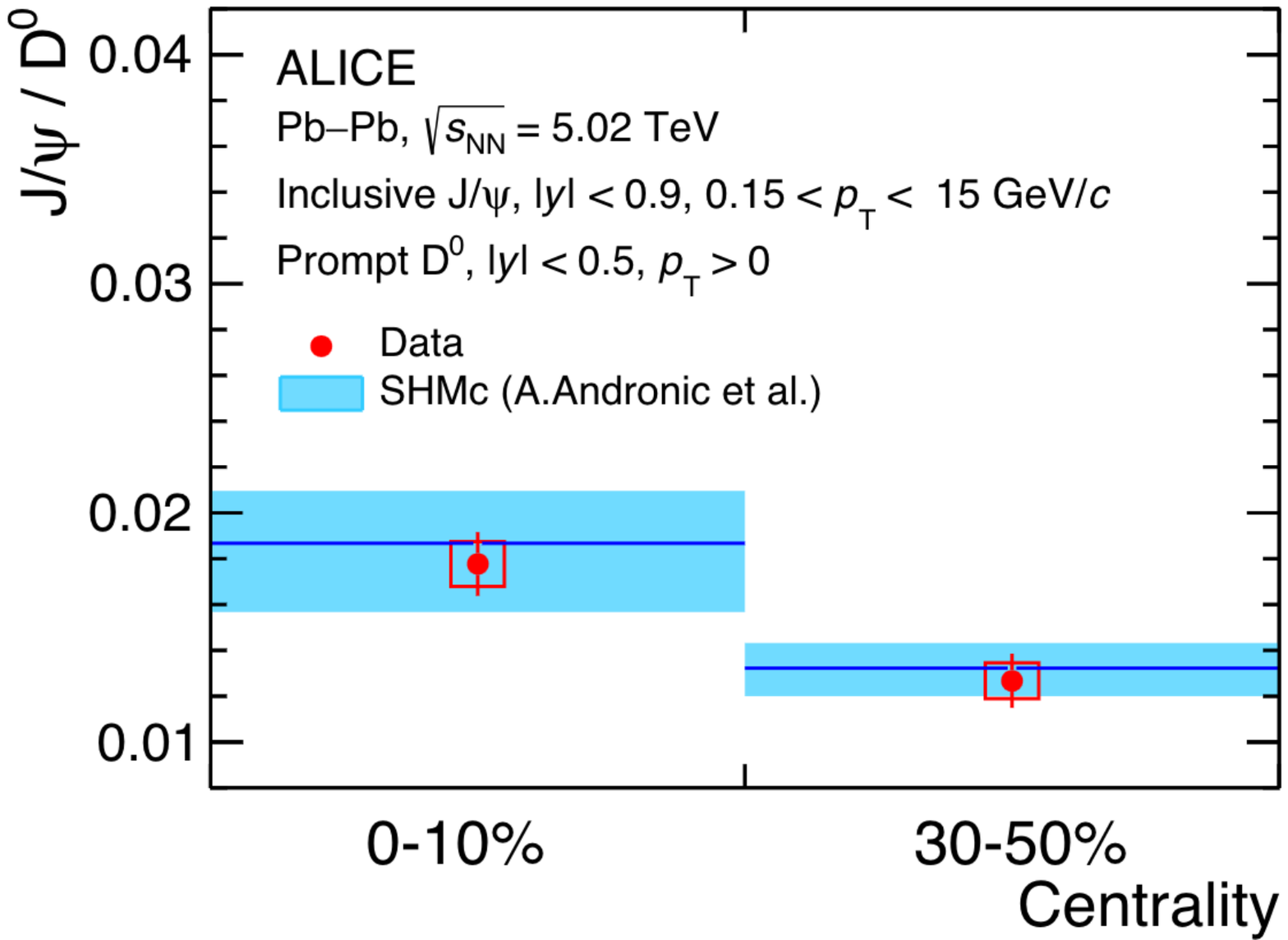}
    \caption{ J/$\psi$ to $D^{0}$ ratio measured in Pb--Pb collisions at the LHC and predicted by the Statistical Hadronization Model with charm SHMc. Figure from~\cite{ALICE:2023gco}.
    }
    \label{fig:JpsitoD0}
\end{figure}

For the description of yields of charmonia, feeding from excited charmonia is very small because of their strong Boltzmann suppression. For open charm mesons and baryons, this is not the case and feeding from excited $D^*$, $\Lambda_c^*$, and $\Sigma_c^*$ is an essential ingredient for the description of open charm hadrons ~\cite{Andronic:2021erx}. Even though the experimental mass spectrum of excited open charm hadrons is not complete, in particular in the baryon sector, the prediction of yields for D-mesons and $\Lambda_c$ baryons compares very well with the measurements\footnote{For $\Lambda_c$ baryons one has to augment the currently measured charm baryon spectrum to account for the large number of additional states predicted by LQCD to achieve agreement with experimental data~\cite{Andronic:2023ioz}.}, both concerning transverse momentum and centrality dependence.

A particularly transparent way to look at the data for Pb--Pb collisions is obtained by analyzing the centrality dependence of the yield ratio $(J/\psi)/D^0$ and comparing the results to the predictions of the SHMc. Recently, both the $D^0$ and $J/\psi$ production cross sections have been well measured down to $p_t$ = 0. The yield ratio $(J/\psi)/D_0$ is reproduced with very good precision for both measured centralities, as demonstrated in Fig.~\ref{fig:JpsitoD0}. This result lends strong support to the assumption that open and hidden charm states are both produced by statistical hadronization at the phase boundary. A more extensive comparison between SHMc and data for open charm hadrons is shown in~\cite{Andronic:2021erx,ALICE:2022wpn}.

From the successful comparison of measured yields for the production of (u,d,s) as well as open and hidden charm hadrons obtained from the SHM or SHMc with essentially only the temperature as a free parameter at LHC energies, one may draw a number of important conclusions. 
\begin{itemize}
\item 
First, we note that hadron production in relativistic nuclear collisions  is described quantitatively by the chemical freeze-out parameters  ($T_{CF}, \mu_{B}$). Note that the fireball volume appearing in the partition function is determined by normalization to the measured number of primary charged particles. At least for energies $\sqrt{s_{NN}}\geq$ 10 GeV these freeze-out parameters agree with good precision with the results from LQCD for the location of the chiral cross over transition. Under these conditions, hadronization is independent of particle species and only dependent on the values of $T$ and $\mu_B$ at the phase boundary. At LHC energy, the chemical potentials vanish, and only $T = T_{c}$ is needed to describe hadronization. 
\item
The mechanism implemented in the SHMc for the production of charmed hadrons implies that these particles are produced from uncorrelated, thermalized charm quarks as is expected for a strongly coupled, deconfined QGP (see also the discussion in ~\cite{Andronic:2021erx}).  At LHC energy, where chemical freeze-out takes place  for central Pb--Pb collisions in a volume per unit rapidity of $V \approx 5000 $ fm$^3$, this implies that charm quarks can travel over linear distances of order 10 fm (see ~\cite{Andronic:2017pug,Andronic:2021erx} for more detail). This result provides strong evidence for deconfinement in the charm sector.
Recent theoretical studies provide  some arguments
for the presence of a confined phase at temperatures above the hadron-resonance gas phase but below the QGP phase~\cite{Cohen:2023hbq,Fujimoto:2025sxx}. The observation of deconfined charm quarks above $T_c$ does not lend support for the existence of such a phase, since colorless charmonium cannot be formed from colored charm and anti-charm quarks in absence of gluons.
\end{itemize}

Future measurement campaigns at the LHC will yield detailed information on the production cross sections of hadrons with multiple charm quarks as well as excited charmonia. The predictions from the SHMc for the relevant cross sections exhibit a rather dramatic hie\-rarchy of enhancements~\cite{Andronic:2021erx} for such processes. Expe\-ri\-men\-tal tests of these predictions would lead to a fundamental understanding of confinement/deconfinement and hadronization. The vision is to obtain, from the measured charmonium spectrum compared to SHMc, a deconfinement temperature similar in spirit to the above cited freeze-out temperature for nuclei.

\section{Some open issues}

We have demonstrated that the SHM and SHMc approaches provide an excellent description of hadron production in relativistic nuclear collisions from  $\sqrt{s_{NN}}= 3$ GeV into the multi-TeV range. Recently there has been increased interest in searching for quark-gluon plasma signatures also in light systems such as pA and pp collisions ~\cite{Grosse-Oetringhaus:2024bwr}. This has led to investigations of SHMc predictions also for these systems ~\cite{Braun-Munzinger:2024ybd} and established new systematics which is rather well visible in the available data, with the exception of the $D^0_s$ meson whose yield is unexpectedly suppressed by about a factor of 2 compared to SHMc predictions. Furthermore, the degree of equilibration reached in the light systems has not yet been well determined.

The detailed production mechanism for composite objects such as light nuclei and hyper-nuclei has not yet been fully established, despite significant experimental efforts. As cases in point we note the system size dependence of the loosely bound hyper-triton which for pp collisions is better understood in a coalescence framework while $^4$He and mass-4 hypernucleus production in heavy systems is well described in the SHM approach.  We note in this context that a possible connection between coalescence models and SHM was already recognized more than 30 years ago~\cite{Baltz:1993jh,Braun-Munzinger:1994zkz}, see also a recent study in~\cite{Reichert:2022mek}. Clearly, very loosely bound objects such as hyper-triton with radius of order of 10 fm cannot survive even the dilute hadronic phase after chemical freeze-out. In ~\cite{Andronic:2017pug} it was then speculated that such objects are first produced as compact (multi-quark) clusters with mass and quantum number of the final state such as hyper-triton and only significantly later evolve into the full nuclear wave function. How this evolution takes place is not however understood. On the other hand, hyper-triton cannot directly coalesce into the fully formed large object as the coalescence process has to take place via the strong interaction which is very short range ($1-2$ fm). So in both models the initial configuration has to be compact. The importance of this aspect is not currently taken into account in the coalescence approach.

A possible way of how to deal in the SHM with the production of objects whose size is comparable to or even much larger than the size of the fireball formed in the collision was recently proposed in a schematic way in ~\cite{Muller:2022uuv}. This mechanism was very recently implemented into our version of the SHM but indeed does not lead to agreement with the measured system-size dependence of the production yields of light nuclei~\cite{Andronic:2026aaa}, where the canonical suppression is sufficient to describe the experimental data.

Finally, we recognize the efforts by Cohen and collaborators ~\cite{Cai:2019jtk,Cohen:2024wgs} to shed light on the 'light nucleus puzzle' but do not share their understanding of the non-equilibrium dilute hadronic phase after chemical freeze-out in SHM.

\vspace{.5cm}
\textbf{Acknowledgements}
K.R. acknowledges support from the National Science Centre (NCN), Poland, under OPUS Grant No. 2022/45/B/ST2/01527, and of the Polish Ministry of Science and Higher Education. This work is part of and supported by the DFG Collaborative Research Centre, SFB1225/\ ISOQUANT.

\bibliography{cracow2026}

\providecommand{\href}[2]{#2}\begingroup\raggedright\begin{thebibliography}{10}

\bibitem{Itoh:1970uw}
N.~Itoh, ``{Hydrostatic Equilibrium of Hypothetical Quark Stars},''
\href{http://dx.doi.org/10.1143/PTP.44.291}{{\em Prog. Theor. Phys.} {\bfseries
  44} (1970) 291}.

\bibitem{Collins:1974ky}
J.~C. Collins and M.~Perry, ``{Superdense Matter: Neutrons Or Asymptotically
  Free Quarks?},''
\href{http://dx.doi.org/10.1103/PhysRevLett.34.1353}{{\em Phys. Rev. Lett.}
  {\bfseries 34} (1975) 1353}.

\bibitem{Cabibbo:1975ig}
N.~Cabibbo and G.~Parisi, ``{Exponential Hadronic Spectrum and Quark
  Liberation},''
\href{http://dx.doi.org/10.1016/0370-2693(75)90158-6}{{\em Phys. Lett. B}
  {\bfseries 59} (1975) 67--69}.

\bibitem{Chapline:1976gy}
G.~Chapline and M.~Nauenberg, ``{Asymptotic Freedom and the Baryon-Quark Phase
  Transition},''
\href{http://dx.doi.org/10.1103/PhysRevD.16.450}{{\em Phys. Rev. D} {\bfseries
  16} (1977) 450}.

\bibitem{Shuryak:1978ij}
E.~V. Shuryak, ``{Quark-gluon plasma and hadronic production of leptons,
  photons and psions},''
\href{http://dx.doi.org/10.1016/0370-2693(78)90370-2}{{\em Phys. Lett. B}
  {\bfseries 78} (1978) 150}.

\bibitem{Boyanovsky:2006bf}
D.~Boyanovsky, H.~de~Vega, and D.~Schwarz, ``{Phase transitions in the early
  and the present universe},''
  \href{http://dx.doi.org/10.1146/annurev.nucl.56.080805.140539}{{\em Ann. Rev.
  Nucl. Part. Sci.} {\bfseries 56} (2006) 441--500},
\href{http://arxiv.org/abs/hep-ph/0602002}{{\ttfamily arXiv:hep-ph/0602002
  [hep-ph]}}.

\bibitem{Busza:2018rrf}
W.~Busza, K.~Rajagopal, and W.~van~der Schee, ``{Heavy Ion Collisions: The Big
  Picture, and the Big Questions},''
  \href{http://dx.doi.org/10.1146/annurev-nucl-101917-020852}{{\em Ann. Rev.
  Nucl. Part. Sci.} {\bfseries 68} (2018) 339--376},
  \href{http://arxiv.org/abs/1802.04801}{{\ttfamily arXiv:1802.04801
  [hep-ph]}}.

\bibitem{Braun-Munzinger:2015hba}
P.~Braun-Munzinger, V.~Koch, T.~Sch{\"a}fer, and J.~Stachel, ``{Properties of
  hot and dense matter from relativistic heavy ion collisions},''
  \href{http://dx.doi.org/10.1016/j.physrep.2015.12.003}{{\em Phys. Rept.}
  {\bfseries 621} (2016) 76--126},
\href{http://arxiv.org/abs/1510.00442}{{\ttfamily arXiv:1510.00442 [nucl-th]}}.

\bibitem{BraunMunzinger:2003zd}
P.~Braun-Munzinger, K.~Redlich, and J.~Stachel, ``{Particle production in heavy
  ion collisions},'' {\em In Hwa, R.C. and Wang, X.N. (eds.): Quark-Gluon
  Plasma 3} (2003) 491--599,
\href{http://arxiv.org/abs/nucl-th/0304013}{{\ttfamily arXiv:nucl-th/0304013
  [nucl-th]}}.

\bibitem{Andronic:2005yp}
A.~Andronic, P.~Braun-Munzinger, and J.~Stachel, ``{Hadron production in
  central nucleus-nucleus collisions at chemical freeze-out},''
  \href{http://dx.doi.org/10.1016/j.nuclphysa.2006.03.012}{{\em Nucl. Phys. A}
  {\bfseries 772} (2006) 167--199},
  \href{http://arxiv.org/abs/nucl-th/0511071}{{\ttfamily
  arXiv:nucl-th/0511071}}.

\bibitem{Andronic:2017pug}
A.~Andronic, P.~Braun-Munzinger, K.~Redlich, and J.~Stachel, ``{Decoding the
  phase structure of QCD via particle production at high energy},''
  \href{http://dx.doi.org/10.1038/s41586-018-0491-6}{{\em Nature} {\bfseries
  561} no.~7723, (2018) 321--330},
  \href{http://arxiv.org/abs/1710.09425}{{\ttfamily arXiv:1710.09425
  [nucl-th]}}.

\bibitem{Bazavov:2018mes}
{\bfseries HotQCD} Collaboration, A.~Bazavov {\em et~al.}, ``{Chiral crossover
  in QCD at zero and non-zero chemical potentials},''
  \href{http://dx.doi.org/10.1016/j.physletb.2019.05.013}{{\em Phys. Lett. B}
  {\bfseries 795} (2019) 15--21},
  \href{http://arxiv.org/abs/1812.08235}{{\ttfamily arXiv:1812.08235
  [hep-lat]}}.

\bibitem{Borsanyi:2020fev}
S.~Borsanyi, Z.~Fodor, J.~N. Guenther, R.~Kara, S.~D. Katz, P.~Parotto,
  A.~Pasztor, C.~Ratti, and K.~K. Szabo, ``{QCD Crossover at Finite Chemical
  Potential from Lattice Simulations},''
  \href{http://dx.doi.org/10.1103/PhysRevLett.125.052001}{{\em Phys. Rev.
  Lett.} {\bfseries 125} no.~5, (2020) 052001},
  \href{http://arxiv.org/abs/2002.02821}{{\ttfamily arXiv:2002.02821
  [hep-lat]}}.

\bibitem{Bonati:2018nut}
C.~Bonati, M.~D'Elia, F.~Negro, F.~Sanfilippo, and K.~Zambello, ``{Curvature of
  the pseudocritical line in QCD: Taylor expansion matches analytic
  continuation},'' \href{http://dx.doi.org/10.1103/PhysRevD.98.054510}{{\em
  Phys. Rev. D} {\bfseries 98} no.~5, (2018) 054510},
  \href{http://arxiv.org/abs/1805.02960}{{\ttfamily arXiv:1805.02960
  [hep-lat]}}.

\bibitem{Aoki:2006we}
Y.~Aoki, G.~Endrodi, Z.~Fodor, S.~Katz, and K.~Szabo, ``{The order of the
  quantum chromodynamics transition predicted by the standard model of particle
  physics},'' \href{http://dx.doi.org/10.1038/nature05120}{{\em Nature}
  {\bfseries 443} (2006) 675--678},
\href{http://arxiv.org/abs/hep-lat/0611014}{{\ttfamily arXiv:hep-lat/0611014
  [hep-lat]}}.

\bibitem{Borsanyi:2024xrx}
S.~Borsanyi, Z.~Fodor, J.~N. Guenther, P.~Parotto, A.~Pasztor, L.~Pirelli,
  K.~K. Szabo, and C.~H. Wong, ``{QCD deconfinement transition line up to
  {\ensuremath{\mu}}B=400{\,}{\,}MeV from finite volume lattice simulations},''
  \href{http://dx.doi.org/10.1103/PhysRevD.110.114507}{{\em Phys. Rev. D}
  {\bfseries 110} no.~11, (2024) 114507},
  \href{http://arxiv.org/abs/2410.06216}{{\ttfamily arXiv:2410.06216
  [hep-lat]}}.

\bibitem{Fischer:2026uni}
C.~S. Fischer and J.~M. Pawlowski, ``{Phase structure and observables at high
  densities from first principles QCD},''
  \href{http://arxiv.org/abs/2603.11135}{{\ttfamily arXiv:2603.11135
  [hep-ph]}}.

\bibitem{Braun-Munzinger:2026krf}
P.~Braun-Munzinger, A.~Rustamov, and N.~Xu, ``{The phase structure of QCD:
  Fluctuations and Correlations},''
  \href{http://arxiv.org/abs/2601.18666}{{\ttfamily arXiv:2601.18666
  [nucl-ex]}}.

\bibitem{Bazavov:2017dus}
A.~Bazavov {\em et~al.}, ``{The QCD Equation of State to $\mathcal{O}(\mu_B^6)$
  from Lattice QCD},'' \href{http://dx.doi.org/10.1103/PhysRevD.95.054504}{{\em
  Phys. Rev. D} {\bfseries 95} no.~5, (2017) 054504},
\href{http://arxiv.org/abs/1701.04325}{{\ttfamily arXiv:1701.04325 [hep-lat]}}.

\bibitem{Zyla:2020zbs}
{\bfseries Particle Data Group} Collaboration, P.~Zyla {\em et~al.}, ``{Review
  of Particle Physics},'' \href{http://dx.doi.org/10.1093/ptep/ptaa104}{{\em
  PTEP} {\bfseries 2020} no.~8, (2020) 083C01}.

\bibitem{Andronic:2018qqt}
A.~Andronic, P.~Braun-Munzinger, B.~Friman, P.~M. Lo, K.~Redlich, and
  J.~Stachel, ``{The thermal proton yield anomaly in Pb-Pb collisions at the
  LHC and its resolution},''
  \href{http://dx.doi.org/10.1016/j.physletb.2019.03.052}{{\em Phys. Lett. B}
  {\bfseries 792} (2019) 304--309},
  \href{http://arxiv.org/abs/1808.03102}{{\ttfamily arXiv:1808.03102
  [hep-ph]}}.

\bibitem{Cleymans:2020fsc}
J.~Cleymans, P.~M. Lo, K.~Redlich, and N.~Sharma, ``{Multiplicity dependence of
  (multi)strange baryons in the canonical ensemble with phase shift
  corrections},'' \href{http://dx.doi.org/10.1103/PhysRevC.103.014904}{{\em
  Phys. Rev. C} {\bfseries 103} no.~1, (2021) 014904},
  \href{http://arxiv.org/abs/2009.04844}{{\ttfamily arXiv:2009.04844
  [hep-ph]}}.

\bibitem{Yasui:2026vve}
S.~Yasui, S.~H. Lee, P.~M. Lo, and C.~Sasaki, ``{New nonet scalar mesons and
  glueballs: the mass spectra and the production yields in relativistic heavy
  ion collisions},'' \href{http://arxiv.org/abs/2603.13764}{{\ttfamily
  arXiv:2603.13764 [hep-ph]}}.

\bibitem{Acharya:2017bso}
{\bfseries ALICE} Collaboration, S.~Acharya {\em et~al.}, ``{Production of
  $^{4}$He and $^{4}\overline{\textrm{He}}$ in Pb-Pb collisions at
  $\sqrt{s_{\mathrm{NN}}}$ = 2.76 TeV at the LHC},''
  \href{http://dx.doi.org/10.1016/j.nuclphysa.2017.12.004}{{\em Nucl. Phys. A}
  {\bfseries 971} (2018) 1--20},
\href{http://arxiv.org/abs/1710.07531}{{\ttfamily arXiv:1710.07531 [nucl-ex]}}.

\bibitem{ALICE:2023ulv}
{\bfseries ALICE} Collaboration, S.~Acharya {\em et~al.}, ``{Measurements of
  Chemical Potentials in Pb-Pb Collisions at sNN=5.02{\,}{\,}TeV},''
  \href{http://dx.doi.org/10.1103/PhysRevLett.133.092301}{{\em Phys. Rev.
  Lett.} {\bfseries 133} no.~9, (2024) 092301},
  \href{http://arxiv.org/abs/2311.13332}{{\ttfamily arXiv:2311.13332
  [nucl-ex]}}.

\bibitem{Andronic:2021dkw}
A.~Andronic, P.~Braun-Munzinger, K.~Redlich, and J.~Stachel, ``{Hadron yields
  in central nucleus-nucleus collisions, the statistical hadronization model
  and the QCD phase diagram},'' in {\em {Criticality in QCD and the Hadron
  Resonance Gas}}.
\newblock 1, 2021.
\newblock \href{http://arxiv.org/abs/2101.05747}{{\ttfamily arXiv:2101.05747
  [nucl-th]}}.

\bibitem{Andronic:2020iyg}
A.~Andronic, P.~Braun-Munzinger, D.~G{\"u}nd{\"u}z, Y.~Kirchhoff, M.~K.
  K{\"o}hler, J.~Stachel, and M.~Winn, ``{Influence of modified light-flavor
  hadron spectra on particle yields in the statistical hadronization model},''
  \href{http://dx.doi.org/10.1016/j.nuclphysa.2021.122176}{{\em Nucl. Phys. A}
  {\bfseries 1010} (2021) 122176},
  \href{http://arxiv.org/abs/2011.03826}{{\ttfamily arXiv:2011.03826
  [nucl-th]}}.

\bibitem{Andronic:2010qu}
A.~Andronic, P.~Braun-Munzinger, J.~Stachel, and H.~St{\"o}cker, ``{Production
  of light nuclei, hypernuclei and their antiparticles in relativistic nuclear
  collisions},'' \href{http://dx.doi.org/10.1016/j.physletb.2011.01.053}{{\em
  Phys. Lett. B} {\bfseries 697} (2011) 203--207},
\href{http://arxiv.org/abs/1010.2995}{{\ttfamily arXiv:1010.2995 [nucl-th]}}.

\bibitem{Braun-Munzinger:2018hat}
P.~Braun-Munzinger and B.~D\"onigus, ``{Loosely-bound objects produced in
  nuclear collisions at the LHC},''
  \href{http://dx.doi.org/10.1016/j.nuclphysa.2019.02.006}{{\em Nucl. Phys. A}
  {\bfseries 987} (2019) 144--201},
  \href{http://arxiv.org/abs/1809.04681}{{\ttfamily arXiv:1809.04681
  [nucl-ex]}}.

\bibitem{Andronic:2014zha}
A.~Andronic, ``{An overview of the experimental study of quark-gluon matter in
  high-energy nucleus-nucleus collisions},''
  \href{http://dx.doi.org/10.1142/S0217751X14300476}{{\em Int. J. Mod. Phys. A}
  {\bfseries 29} (2014) 1430047},
\href{http://arxiv.org/abs/1407.5003}{{\ttfamily arXiv:1407.5003 [nucl-ex]}}.

\bibitem{Adamczyk:2017iwn}
{\bfseries STAR} Collaboration, L.~Adamczyk {\em et~al.}, ``{Bulk Properties of
  the Medium Produced in Relativistic Heavy-Ion Collisions from the Beam Energy
  Scan Program},'' \href{http://dx.doi.org/10.1103/PhysRevC.96.044904}{{\em
  Phys. Rev. C} {\bfseries 96} no.~4, (2017) 044904},
\href{http://arxiv.org/abs/1701.07065}{{\ttfamily arXiv:1701.07065 [nucl-ex]}}.

\bibitem{Hagedorn:1984uy}
R.~Hagedorn and K.~Redlich, ``{Statistical Thermodynamics in Relativistic
  Particle and Ion Physics: Canonical or Grand Canonical?},''
\href{http://dx.doi.org/10.1007/BF01436508}{{\em Z. Phys. C} {\bfseries 27}
  (1985) 541}.

\bibitem{Hamieh:2000tk}
S.~Hamieh, K.~Redlich, and A.~Tounsi, ``{Canonical description of strangeness
  enhancement from p-A to Pb Pb collisions},''
  \href{http://dx.doi.org/10.1016/S0370-2693(00)00762-0}{{\em Phys. Lett.}
  {\bfseries B486} (2000) 61--66},
\href{http://arxiv.org/abs/hep-ph/0006024}{{\ttfamily arXiv:hep-ph/0006024
  [hep-ph]}}.

\bibitem{Braun-Munzinger:2003pwq}
P.~Braun-Munzinger, K.~Redlich, and J.~Stachel, ``{Particle production in heavy
  ion collisions. In {\it Quark Gluon Plasma 3}, eds. R. C. Hwa and Xin-Nian
  Wang, World Scientific Publishing, 491--599},''
  \href{http://arxiv.org/abs/nucl-th/0304013}{{\ttfamily
  arXiv:nucl-th/0304013}}.

\bibitem{BraunMunzinger:1998cg}
P.~Braun-Munzinger and J.~Stachel, ``{Dynamics of ultrarelativistic nuclear
  collisions with heavy beams: An Experimental overview},''
  \href{http://dx.doi.org/10.1016/S0375-9474(98)00342-X}{{\em Nucl. Phys. A}
  {\bfseries 638} (1998) 3--18},
\href{http://arxiv.org/abs/nucl-ex/9803015}{{\ttfamily arXiv:nucl-ex/9803015
  [nucl-ex]}}.

\bibitem{Stock:1999hm}
R.~Stock, ``{The parton to hadron phase transition observed in Pb+Pb collisions
  at 158-GeV per nucleon},''
  \href{http://dx.doi.org/10.1016/S0370-2693(99)00482-7}{{\em Phys. Lett. B}
  {\bfseries 456} (1999) 277--282},
\href{http://arxiv.org/abs/hep-ph/9905247}{{\ttfamily arXiv:hep-ph/9905247
  [hep-ph]}}.

\bibitem{BraunMunzinger:2003zz}
P.~Braun-Munzinger, J.~Stachel, and C.~Wetterich, ``{Chemical freezeout and the
  QCD phase transition temperature},''
  \href{http://dx.doi.org/10.1016/j.physletb.2004.05.081}{{\em Phys. Lett. B}
  {\bfseries 596} (2004) 61--69},
\href{http://arxiv.org/abs/nucl-th/0311005}{{\ttfamily arXiv:nucl-th/0311005
  [nucl-th]}}.

\bibitem{Andronic:2008gu}
A.~Andronic, P.~Braun-Munzinger, and J.~Stachel, ``{Thermal hadron production
  in relativistic nuclear collisions: the hadron mass spectrum, the horn, and
  the QCD phase transition},''
  \href{http://dx.doi.org/10.1016/j.physletb.2009.06.021,
  10.1016/j.physletb.2009.02.014}{{\em Phys. Lett. B} {\bfseries 673} (2009)
  142--145},
\href{http://arxiv.org/abs/0812.1186}{{\ttfamily arXiv:0812.1186 [nucl-th]}}.

\bibitem{Hagedorn:1965st}
R.~Hagedorn, ``{Statistical thermodynamics of strong interactions at
  high-energies},''
{\em Nuovo Cim. Suppl.} {\bfseries 3} (1965) 147--186.

\bibitem{Cleymans:1998yb}
J.~Cleymans, H.~Oeschler, and K.~Redlich, ``{Influence of impact parameter on
  thermal description of relativistic heavy ion collisions at (1-2) A-GeV},''
  \href{http://dx.doi.org/10.1103/PhysRevC.59.1663}{{\em Phys. Rev. C}
  {\bfseries 59} (1999) 1663},
\href{http://arxiv.org/abs/nucl-th/9809027}{{\ttfamily arXiv:nucl-th/9809027
  [nucl-th]}}.

\bibitem{Vovchenko:2015idt}
V.~Vovchenko, V.~V. Begun, and M.~I. Gorenstein, ``{Hadron multiplicities and
  chemical freeze-out conditions in proton-proton and nucleus-nucleus
  collisions},'' \href{http://dx.doi.org/10.1103/PhysRevC.93.064906}{{\em Phys.
  Rev. C} {\bfseries 93} no.~6, (2016) 064906},
\href{http://arxiv.org/abs/1512.08025}{{\ttfamily arXiv:1512.08025 [nucl-th]}}.

\bibitem{Becattini:2016xct}
F.~Becattini, J.~Steinheimer, R.~Stock, and M.~Bleicher, ``{Hadronization
  conditions in relativistic nuclear collisions and the QCD pseudo-critical
  line},'' \href{http://dx.doi.org/10.1016/j.physletb.2016.11.033}{{\em Phys.
  Lett. B} {\bfseries 764} (2017) 241--246},
\href{http://arxiv.org/abs/1605.09694}{{\ttfamily arXiv:1605.09694 [nucl-th]}}.

\bibitem{ALICE:2021rxa}
{\bfseries ALICE} Collaboration, S.~Acharya {\em et~al.}, ``{Prompt D$^{0}$,
  D$^{+}$, and D$^{*+}$ production in Pb{\textendash}Pb collisions at $
  \sqrt{s_{\mathrm{NN}}} $ = 5.02 TeV},''
  \href{http://dx.doi.org/10.1007/JHEP01(2022)174}{{\em JHEP} {\bfseries 01}
  (2022) 174}, \href{http://arxiv.org/abs/2110.09420}{{\ttfamily
  arXiv:2110.09420 [nucl-ex]}}.

\bibitem{Andronic:2021erx}
A.~Andronic, P.~Braun-Munzinger, M.~K. K{\"o}hler, A.~Mazeliauskas, K.~Redlich,
  J.~Stachel, and V.~Vislavicius, ``{The multiple-charm hierarchy in the
  statistical hadronization model},''
  \href{http://dx.doi.org/10.1007/JHEP07(2021)035}{{\em JHEP} {\bfseries 07}
  (2021) 035}, \href{http://arxiv.org/abs/2104.12754}{{\ttfamily
  arXiv:2104.12754 [hep-ph]}}.

\bibitem{Andronic:2019wva}
A.~Andronic, P.~Braun-Munzinger, M.~K. K\"ohler, K.~Redlich, and J.~Stachel,
  ``{Transverse momentum distributions of charmonium states with the
  statistical hadronization model},''
  \href{http://dx.doi.org/10.1016/j.physletb.2019.134836}{{\em Phys. Lett. B}
  {\bfseries 797} (2019) 134836},
  \href{http://arxiv.org/abs/1901.09200}{{\ttfamily arXiv:1901.09200
  [nucl-th]}}.

\bibitem{Altenkort:2020fgs}
L.~Altenkort, A.~M. Eller, O.~Kaczmarek, L.~Mazur, G.~D. Moore, and H.-T. Shu,
  ``{Heavy quark momentum diffusion from the lattice using gradient flow},''
  \href{http://dx.doi.org/10.1103/PhysRevD.103.014511}{{\em Phys. Rev. D}
  {\bfseries 103} no.~1, (2021) 014511},
  \href{http://arxiv.org/abs/2009.13553}{{\ttfamily arXiv:2009.13553
  [hep-lat]}}.

\bibitem{Altenkort:2023oms}
{\bfseries HotQCD} Collaboration, L.~Altenkort, O.~Kaczmarek, R.~Larsen,
  S.~Mukherjee, P.~Petreczky, H.-T. Shu, and S.~Stendebach, ``{Heavy Quark
  Diffusion from 2+1 Flavor Lattice QCD with 320~MeV Pion Mass},''
  \href{http://dx.doi.org/10.1103/PhysRevLett.130.231902}{{\em Phys. Rev.
  Lett.} {\bfseries 130} no.~23, (2023) 231902},
  \href{http://arxiv.org/abs/2302.08501}{{\ttfamily arXiv:2302.08501
  [hep-lat]}}.

\bibitem{ALICE:2013xna}
{\bfseries ALICE} Collaboration, E.~Abbas {\em et~al.}, ``{J/$\psi$ elliptic
  flow in Pb--Pb Collisions at $\sqrt{s_{_{\rm NN}}}$=2.76 TeV},''
  \href{http://dx.doi.org/10.1103/PhysRevLett.111.162301}{{\em Phys. Rev.
  Lett.} {\bfseries 111} (2013) 162301},
\href{http://arxiv.org/abs/1303.5880}{{\ttfamily arXiv:1303.5880 [nucl-ex]}}.

\bibitem{He:2021zej}
M.~He, B.~Wu, and R.~Rapp, ``{Collectivity of J/{\ensuremath{\psi}} Mesons in
  Heavy-Ion Collisions},''
  \href{http://dx.doi.org/10.1103/PhysRevLett.128.162301}{{\em Phys. Rev.
  Lett.} {\bfseries 128} no.~16, (2022) 162301},
  \href{http://arxiv.org/abs/2111.13528}{{\ttfamily arXiv:2111.13528
  [nucl-th]}}.

\bibitem{Cho:2019lxb}
S.~Cho, K.-J. Sun, C.~M. Ko, S.~H. Lee, and Y.~Oh, ``{Charmed hadron production
  in an improved quark coalescence model},''
  \href{http://dx.doi.org/10.1103/PhysRevC.101.024909}{{\em Phys. Rev. C}
  {\bfseries 101} no.~2, (2020) 024909},
  \href{http://arxiv.org/abs/1905.09774}{{\ttfamily arXiv:1905.09774
  [nucl-th]}}.

\bibitem{ExHIC:2017smd}
{\bfseries ExHIC} Collaboration, S.~Cho {\em et~al.}, ``{Exotic hadrons from
  heavy ion collisions},''
  \href{http://dx.doi.org/10.1016/j.ppnp.2017.02.002}{{\em Prog. Part. Nucl.
  Phys.} {\bfseries 95} (2017) 279--322},
  \href{http://arxiv.org/abs/1702.00486}{{\ttfamily arXiv:1702.00486
  [nucl-th]}}.

\bibitem{Zhou:2014kka}
K.~Zhou, N.~Xu, Z.~Xu, and P.~Zhuang, ``{Medium effects on charmonium
  production at ultrarelativistic energies available at the CERN Large Hadron
  Collider},'' \href{http://dx.doi.org/10.1103/PhysRevC.89.054911}{{\em Phys.
  Rev. C} {\bfseries 89} (2014) 054911},
\href{http://arxiv.org/abs/1401.5845}{{\ttfamily arXiv:1401.5845 [nucl-th]}}.

\bibitem{Greco:2003vf}
V.~Greco, C.~Ko, and R.~Rapp, ``{Quark coalescence for charmed mesons in
  ultrarelativistic heavy ion collisions},''
  \href{http://dx.doi.org/10.1016/j.physletb.2004.06.064}{{\em Phys. Lett. B}
  {\bfseries 595} (2004) 202--208},
\href{http://arxiv.org/abs/nucl-th/0312100}{{\ttfamily arXiv:nucl-th/0312100
  [nucl-th]}}.

\bibitem{Aarts:2016hap}
G.~Aarts {\em et~al.}, ``{Heavy-flavor production and medium properties in
  high-energy nuclear collisions - What next?},''
  \href{http://dx.doi.org/10.1140/epja/i2017-12282-9}{{\em Eur. Phys. J. A}
  {\bfseries 53} no.~5, (2017) 93},
  \href{http://arxiv.org/abs/1612.08032}{{\ttfamily arXiv:1612.08032
  [nucl-th]}}.

\bibitem{Maiani:2022psl}
L.~Maiani and A.~Pilloni, ``{GGI Lectures on Exotic Hadrons},''
\newblock 7, 2022.
\newblock \href{http://arxiv.org/abs/2207.05141}{{\ttfamily arXiv:2207.05141
  [hep-ph]}}.

\bibitem{Song:2021mvc}
T.~Song and G.~Coci, ``{Prerequisites for heavy quark coalescence in heavy-ion
  collisions},'' \href{http://dx.doi.org/10.1016/j.nuclphysa.2022.122539}{{\em
  Nucl. Phys. A} {\bfseries 1028} (2022) 122539},
  \href{http://arxiv.org/abs/2104.10987}{{\ttfamily arXiv:2104.10987
  [nucl-th]}}.

\bibitem{Braun-Munzinger:2000csl}
P.~Braun-Munzinger and J.~Stachel, ``{(Non)thermal aspects of charmonium
  production and a new look at J / psi suppression},''
  \href{http://dx.doi.org/10.1016/S0370-2693(00)00991-6}{{\em Phys. Lett. B}
  {\bfseries 490} (2000) 196--202},
  \href{http://arxiv.org/abs/nucl-th/0007059}{{\ttfamily
  arXiv:nucl-th/0007059}}.

\bibitem{Andronic:2003zv}
A.~Andronic, P.~Braun-Munzinger, K.~Redlich, and J.~Stachel, ``{Statistical
  hadronization of charm in heavy ion collisions at SPS, RHIC and LHC},''
  \href{http://dx.doi.org/10.1016/j.physletb.2003.07.066}{{\em Phys. Lett. B}
  {\bfseries 571} (2003) 36--44},
\href{http://arxiv.org/abs/nucl-th/0303036}{{\ttfamily arXiv:nucl-th/0303036
  [nucl-th]}}.

\bibitem{Andronic:2006ky}
A.~Andronic, P.~Braun-Munzinger, K.~Redlich, and J.~Stachel, ``{Statistical
  hadronization of heavy quarks in ultra-relativistic nucleus-nucleus
  collisions},'' \href{http://dx.doi.org/10.1016/j.nuclphysa.2007.02.013}{{\em
  Nucl. Phys. A} {\bfseries 789} (2007) 334--356},
\href{http://arxiv.org/abs/nucl-th/0611023}{{\ttfamily arXiv:nucl-th/0611023
  [nucl-th]}}.

\bibitem{Matsui:1986dk}
T.~Matsui and H.~Satz, ``{J/$\psi$ suppression by quark-gluon plasma
  formation},''
\href{http://dx.doi.org/10.1016/0370-2693(86)91404-8}{{\em Phys. Lett. B}
  {\bfseries 178} (1986) 416}.

\bibitem{ALICE:2023gco}
{\bfseries ALICE} Collaboration, S.~Acharya {\em et~al.}, ``{Measurements of
  inclusive J/{\ensuremath{\psi}} production at midrapidity and forward
  rapidity in Pb{\textendash}Pb collisions at sNN = 5.02 TeV},''
  \href{http://dx.doi.org/10.1016/j.physletb.2024.138451}{{\em Phys. Lett. B}
  {\bfseries 849} (2024) 138451},
  \href{http://arxiv.org/abs/2303.13361}{{\ttfamily arXiv:2303.13361
  [nucl-ex]}}.

\bibitem{Andronic:2023ioz}
A.~Andronic, P.~Braun-Munzinger, H.~Brun{\ss}en, J.~Crkovsk{\'a}, J.~Stachel,
  V.~Vislavicius, and M.~V{\"o}lkl, ``{Transverse dynamics of charmed hadrons
  in ultra-relativistic nuclear collisions},''
  \href{http://dx.doi.org/10.1007/JHEP10(2024)229}{{\em JHEP} {\bfseries 10}
  (2024) 229}, \href{http://arxiv.org/abs/2308.14821}{{\ttfamily
  arXiv:2308.14821 [hep-ph]}}.

\bibitem{ALICE:2022wpn}
{\bfseries ALICE} Collaboration, S.~Acharya {\em et~al.}, ``{The ALICE
  experiment: a journey through QCD},''
  \href{http://dx.doi.org/10.1140/epjc/s10052-024-12935-y}{{\em Eur. Phys. J.
  C} {\bfseries 84} no.~8, (2024) 813},
  \href{http://arxiv.org/abs/2211.04384}{{\ttfamily arXiv:2211.04384
  [nucl-ex]}}.

\bibitem{Cohen:2023hbq}
T.~D. Cohen and L.~Y. Glozman, ``{Large $N_c$ QCD phase diagram at $\mu
  _B=0$},'' \href{http://dx.doi.org/10.1140/epja/s10050-024-01400-9}{{\em Eur.
  Phys. J. A} {\bfseries 60} no.~9, (2024) 171},
  \href{http://arxiv.org/abs/2311.07333}{{\ttfamily arXiv:2311.07333
  [hep-ph]}}.

\bibitem{Fujimoto:2025sxx}
Y.~Fujimoto, K.~Fukushima, Y.~Hidaka, and L.~McLerran, ``{New state of matter
  between the hadronic phase and the quark-gluon plasma?},''
  \href{http://dx.doi.org/10.1103/h71y-km92}{{\em Phys. Rev. D} {\bfseries 112}
  no.~7, (2025) 074006}, \href{http://arxiv.org/abs/2506.00237}{{\ttfamily
  arXiv:2506.00237 [hep-ph]}}.

\bibitem{Grosse-Oetringhaus:2024bwr}
J.~F. Grosse-Oetringhaus and U.~A. Wiedemann, ``{A Decade of Collectivity in
  Small Systems},'' \href{http://arxiv.org/abs/2407.07484}{{\ttfamily
  arXiv:2407.07484 [hep-ex]}}.

\bibitem{Braun-Munzinger:2024ybd}
P.~Braun-Munzinger, K.~Redlich, N.~Sharma, and J.~Stachel, ``{Emergence of new
  systematics for open charm production in high energy collisions},''
  \href{http://dx.doi.org/10.1007/JHEP04(2025)058}{{\em JHEP} {\bfseries 04}
  (2025) 058}, \href{http://arxiv.org/abs/2408.07496}{{\ttfamily
  arXiv:2408.07496 [hep-ph]}}.

\bibitem{Baltz:1993jh}
A.~J. Baltz, C.~B. Dover, S.~H. Kahana, Y.~Pang, T.~J. Schlagel, and
  E.~Schnedermann, ``{Strange cluster formation in relativistic heavy ion
  collisions},'' \href{http://dx.doi.org/10.1016/0370-2693(94)90063-9}{{\em
  Phys. Lett. B} {\bfseries 325} (1994) 7--12}.

\bibitem{Braun-Munzinger:1994zkz}
P.~Braun-Munzinger and J.~Stachel, ``{Production of strange clusters and
  strange matter in nucleus-nucleus collisions at the AGS},''
  \href{http://dx.doi.org/10.1088/0954-3899/21/3/002}{{\em J. Phys. G}
  {\bfseries 21} (1995) L17--L20},
  \href{http://arxiv.org/abs/nucl-th/9412035}{{\ttfamily
  arXiv:nucl-th/9412035}}.

\bibitem{Reichert:2022mek}
T.~Reichert, J.~Steinheimer, V.~Vovchenko, B.~D{\"o}nigus, and M.~Bleicher,
  ``{Energy dependence of light hypernuclei production in heavy-ion collisions
  from a coalescence and statistical-thermal model perspective},''
  \href{http://dx.doi.org/10.1103/PhysRevC.107.014912}{{\em Phys. Rev. C}
  {\bfseries 107} no.~1, (2023) 014912},
  \href{http://arxiv.org/abs/2210.11876}{{\ttfamily arXiv:2210.11876
  [nucl-th]}}.

\bibitem{Muller:2022uuv}
B.~M{\"u}ller, ``{SQM2022: Theoretical Summary},''
  \href{http://dx.doi.org/10.1051/epjconf/202327606017}{{\em EPJ Web Conf.}
  {\bfseries 276} (2023) 06017},
  \href{http://arxiv.org/abs/2209.00070}{{\ttfamily arXiv:2209.00070
  [hep-ph]}}.

\bibitem{Andronic:2026aaa}
A.~Andronic, P.~Braun-Munzinger, H.~{Brun\ss en}, and J.~Stachel, ``{The Effect
  of Nuclear Size on the Production of Nuclei and Hypernuclei in the
  Statistical Hadronization Model},'' {\em in preparation} (2026) .

\bibitem{Cai:2019jtk}
Y.~Cai, T.~D. Cohen, B.~A. Gelman, and Y.~Yamauchi, ``{Yields of weakly-bound
  light nuclei as a probe of the statistical hadronization model},''
  \href{http://dx.doi.org/10.1103/PhysRevC.100.024911}{{\em Phys. Rev. C}
  {\bfseries 100} no.~2, (2019) 024911},
  \href{http://arxiv.org/abs/1905.02753}{{\ttfamily arXiv:1905.02753
  [nucl-th]}}.

\bibitem{Cohen:2024wgs}
T.~Cohen and M.~Pradeep, ``{Hypertriton puzzle in relativistic heavy-ion
  collisions},'' \href{http://dx.doi.org/10.1103/PhysRevC.111.054917}{{\em
  Phys. Rev. C} {\bfseries 111} no.~5, (2025) 054917},
  \href{http://arxiv.org/abs/2410.05569}{{\ttfamily arXiv:2410.05569
  [nucl-th]}}.

\end{thebibliography}\endgroup

\end{document}